\documentclass[prd,twocolumn,showpacs,amsmath,amssymb]{revtex4-1}
\usepackage{amssymb}
\usepackage{amsfonts}
\usepackage{overpic,graphicx}
\usepackage{keyval,graphicx}
\usepackage{textcomp,wasysym}
\usepackage[perpage,symbol]{footmisc}
\usepackage[dvipdfm,CJKbookmarks=true, colorlinks=true,
linkcolor=blue, urlcolor=blue,citecolor=blue]{hyperref}
\usepackage{bm}
\usepackage{color}
\usepackage{pstricks}
\usepackage{pst-node}
\usepackage{rotating}
\usepackage{times}
\usepackage{overpic}
\usepackage{ulem,multirow}
\usepackage[english]{babel}
\usepackage{slashbox}
\begin{document}

\title{\boldmath Observation of $\chi_{c2}\to\eta'\eta'$ and $\chi_{c0,2}\to\eta\eta'$}

\author{
\begin{small}
\begin{center}
M.~Ablikim$^{1}$, M.~N.~Achasov$^{9,d}$, S. ~Ahmed$^{14}$, M.~Albrecht$^{4}$, A.~Amoroso$^{53A,53C}$, F.~F.~An$^{1}$, Q.~An$^{50,40}$, J.~Z.~Bai$^{1}$, Y.~Bai$^{39}$, O.~Bakina$^{24}$
% ...
\\
\vspace{0.2cm}
(BESIII Collaboration)\\
\vspace{0.2cm} {\it
$^{1}$ Institute of High Energy Physics, Beijing 100049, People's Republic of China\\
% ...
$^{4}$ Bochum Ruhr-University, D-44780 Bochum, Germany\\
% ...
$^{9}$ G.I. Budker Institute of Nuclear Physics SB RAS (BINP), Novosibirsk 630090, Russia\\
% ...
$^{14}$ Helmholtz Institute Mainz, Johann-Joachim-Becher-Weg 45, D-55099 Mainz, Germany\\
% ...
$^{24}$ Joint Institute for Nuclear Research, 141980 Dubna, Moscow region, Russia\\
% ...
$^{39}$ Southeast University, Nanjing 211100, People's Republic of China\\
$^{40}$ State Key Laboratory of Particle Detection and Electronics, Beijing 100049, Hefei 230026, People's Republic of China\\
% ...
$^{50}$ University of Science and Technology of China, Hefei 230026, People's Republic of China\\
% ...
$^{53}$ (A)University of Turin, I-10125, Turin, Italy; (B)University of Eastern Piedmont, I-15121, Alessandria, Italy; (C)INFN, I-10125, Turin, Italy\\
% ...
$^{d}$ Also at the Novosibirsk State University, Novosibirsk, 630090, Russia\\
% ...
}\end{center}
\vspace{0.4cm}
\end{small}
}
\affiliation{}

\begin{abstract}
Using a sample of $448.1\times 10^6$ $\psi(3686)$ events collected with the BESIII detector in 2009 and 2012,
we study the decays $\chi_{c0,2}\to$ $\eta'\eta'$ and $\eta\eta'$.
The decays $\chi_{c2}\to\eta'\eta'$, $\chi_{c0}\to\eta\eta'$ and $\chi_{c2}\to\eta\eta'$ are
observed for the first time with statistical significances of $9.6\sigma$, $13.4\sigma$ and $7.5\sigma$,
respectively. The branching fractions are determined to be
$\mathcal{B}(\chi_{c0}\to\eta'\eta') =(2.19\pm0.03\pm0.14)\times10^{-3}$,
$\mathcal{B}(\chi_{c2}\to\eta'\eta') = (4.76\pm0.56\pm0.38)\times10^{-5}$,
$\mathcal{B}(\chi_{c0}\to\eta\eta') = (8.92\pm0.84\pm0.65)\times10^{-5}$
and $\mathcal{B}(\chi_{c2}\to\eta\eta') = (2.27\pm0.43\pm0.25)\times10^{-5}$,
where the first uncertainties are statistical and the second are systematic. The precision for the
measurement of $\mathcal{B}(\chi_{c0}\to\eta'\eta')$ is significantly improved compared to previous measurements.
Based on the measured branching fractions, the role played by the doubly
and singly Okubo-Zweig-Iizuka disconnected transition amplitudes for $\chi_{c0,2}$ decays
into pseudoscalar meson pairs can be clarified.
\end{abstract}


\maketitle
\oddsidemargin -0.2cm
\evensidemargin -0.2cm

\section{introduction}

During the past decades an enormous number of decay channels have been measured for
$J/\psi$ and $\psi(3686)$~\footnote{$\psi(3686)$ denotes the state
called $\psi(2S)$ by PDG.}. It can be attributed to the accumulation of high statistics
of $J/\psi$ and $\psi(3686)$ events which can be accessed directly in $e^+e^-$ annihilations.
As a result, many interesting properties associated with the strong decays of $J/\psi$ and $\psi(3686)$
have been investigated and will advance our knowledge about the strong QCD in the interplay
of perturbative and non-perturbative strong interaction regime. In contrast, little is known
about the $\chi_{cJ}$ ($J=0$, $1$, $2$) decays since they can not be produced directly in
$e^{+}e^{-}$ annihilation due to spin-parity conservation. In Ref.~\cite{zhaozou} it was argued that
the ratio of the decay branching fractions between $J/\psi \to \omega f_0(1710)$ and
$J/\psi \to \phi f_0(1710)$~\cite{pdg} encodes the production mechanisms of light quark
contents via the Okubo-Zweig-Iizuka (OZI) rule violations.  In Refs.~\cite{zhaoq1,zhaoq2}
parametrization schemes were proposed in order to further understand the OZI rule violating
mechanisms in the two-body decays of $\chi_{cJ}$ to $SS$, $PP$ and $VV$ ($S$~=~scalar,
$P$~=~pseudoscalar, $V$~=~vector).  It was shown that apart from the singly OZI (SOZI) disconnected
process, the doubly OZI (DOZI) disconnected process may play a crucial role in the production of
isospin-0 light meson pairs, for instance, in $\chi_{cJ}\to f_0 f_0'$, $\omega\omega$, $\phi\phi$,
$\omega\phi$, $\eta\eta$, $\eta\eta'$ and $\eta'\eta'$. By defining the relative strength $r$ between
the DOZI and SOZI violating amplitudes in addition to several other physical quantities in the SU(3)
flavor basis, insights into the mechanisms for producing light meson pairs in charmonium decays can be gained.

Several $\chi_{c0}\to SS$ decay processes have been previously observed and measured~\cite{chicss},
but no definitive conclusions can yet be drawn. In the $\chi_{cJ}\to VV$ sector, BESIII's results~\cite{vv}
indicate that violation of the OZI rule and SU($3$) flavor symmetry breaking
are significant in $\chi_{c0} \to VV$ decays, but small in $\chi_{c2} \to VV$ decays~\cite{zhaoq1}. Furthermore,
the observation of a small $\chi_{c0}\to\omega\phi$ branching fraction and upper limits on
$\chi_{c2}\to\omega\phi$ imply a small DOZI contribution in $\chi_{c0,2} \to VV$ decays.
As for $\chi_{c0,2}\to PP$ decays, most of them have been well measured except for the
processes with final states containing an $\eta'$ meson. Until now, only the branching fraction of $\chi_{c0}\to\eta'\eta'$
is available with poor precision, while no obvious signals for $\chi_{c2}\to\eta'\eta'$
and $\chi_{c0,2}\to \eta\eta'$ are observed~\cite{pdg}. It is worth noting that according to Eq.~$15$ in
Ref.~\cite{zhaoq1} the calculation of $r$ is more sensitive to the branching fractions of
$\chi_{c0,2} \to\eta'\eta'$ and $\eta\eta'$ than those of $\chi_{c0,2} \to\eta\eta$~\cite{zhaoq1,zhaoq2}.
Therefore, measurements of $\chi_{c0,2} \to\eta'\eta'$ and $\eta\eta'$ are desirable and crucial to
disentangle the roles played by OZI violation in charmonium decay.

In this article, we report measurements of the branching fractions of $\chi_{c0,2}\to\eta'\eta'$ and $\eta\eta'$
based on a data sample of $448.1\times 10^6$ $\psi(3686)$ events~\cite{npsip,npsip1}
collected with the BESIII detector~\cite{bes3} operated at the BEPCII storage ring in
2009 and 2012. The number of $\psi(3686)$ events, determined by measuring inclusive hadronic events,
is $(107.0\pm0.8)\times10^{6}$ for 2009 and  $(341.1\pm 2.1)\times10^{6}$ for 2012.

\section{The BESIII detector and simulation}

The BESIII detector is composed of four sub-detectors:
the main drift chamber (MDC), the time-of-flight counter (TOF), the electromagnetic
calorimeter (EMC) and the muon counter (MUC). There is a superconducting solenoid magnet surrounding
the electromagnetic calorimeter, providing a $1$~Tesla ($0.9$~Tesla during 2012 data taking) magnetic field.
The details of the BESIII detector can be found in Ref.~\cite{bes3}. The
BESIII detector is simulated by the {\sc GEANT4}-based~\cite{geant4} simulation software {\sc BOOST}~\cite{boost},
which includes the geometric and material description of the BESIII detector, the detector response and digitization models,
as well as a record of the detector running conditions and performances. The production of the $\psi(3686)$
resonance is simulated by the Monte Carlo (MC) generator {\sc KKMC}~\cite{kkmc}, in which the effects of
beam energy spread and initial state radiation are considered.  Known decays are generated by
{\sc EVTGEN}~\cite{evtgen} using branching fractions quoted by the particle data group (PDG)~\cite{pdg},
and the remaining unknown decays are generated with {\sc LUNDCHARM}~\cite{lundcharm}.
The transition of $\psi(3686) \to \gamma\chi_{cJ}$ is
assumed to be a pure $E1$ process~\cite{E1}. The subsequent decay $\chi_{c0} \to \eta'\eta'/\eta\eta'$ with
$\eta$ and $\eta'$ decay to the specific final states listed in the following paragraph are
generated by assuming a uniform phase space distribution, while the angular distributions of $\eta$ and
$\eta'$ in $\chi_{c2}$ decays are taken as those of $\pi^{\pm}$ in Ref.~\cite{liuzq}, 
which is the measurement with the highest precision until now.

To increase statistics, two dominant $\eta'$ decay modes, $\eta'\to\gamma\pi^+\pi^-$ and
$\eta'\to\eta\pi^+\pi^-$, are considered,
while the $\eta$ is reconstructed in its prominent decay mode $\eta \to \gamma\gamma$.
Consequently, there are three decay modes in the study of $\chi_{c0,2} \to \eta'\eta'$: both $\eta'$
decay to $\gamma\pi^+\pi^-$
(mode A), both $\eta'$ decay to $\eta\pi^+\pi^-$ (mode B), and one $\eta'$ decays to $\gamma\pi^+\pi^-$ while
the other $\eta'$ decays to $\eta\pi^+\pi^-$ (mode C). Two decay modes are considered for
$\chi_{c0,2}\to\eta\eta'$: $\eta'$ decays to $\gamma\pi^+\pi^-$ (mode I) and to $\eta\pi^+\pi^-$ (mode II).

\section{event selection}

Charged tracks are reconstructed using MDC hits within the
acceptance range of $|\cos\theta|<0.93$, where
$\theta$ is the polar angle with respect to the electron beam direction. They are required to
originate from the interaction region, defined as $R_{xy}<1$~cm and $|V_{z}|<10$~cm, where $R_{xy}$ and $|V_{z}|$
are the distances of closest approach in the $xy$-plane and the $z$ direction, respectively.
All charged tracks are assumed to be pions. The candidate photons are selected using EMC showers.
The photon energy deposited in the EMC is required to be
larger than 25 MeV in the barrel region ($|\cos\theta|<0.8$) or 50 MeV in the
end caps region ($0.86<|\cos\theta|<0.92$). The EMC hit time of the photon candidate must be within the
range $0\leq t \leq 700$~ns from the event start time to suppress electronic noise and energy deposits unrelated to the event.
An $\eta$ candidate is reconstructed from a pair of photons
with an invariant mass $M_{\gamma\gamma}$ satisfying $|M_{\gamma\gamma}-M_{\eta}| < 20$~MeV/$c^2$,
where $M_{\eta}$ is the nominal $\eta$ mass~\cite{pdg}.

A four momentum constrained kinematic fit to the initial beam four momentum,
with an additional mass constraint on $\eta$ candidates, is imposed on the candidate charged tracks and
photons with the proper charged tracks and photons hypothesis, to improve the mass resolution and suppress backgrounds. If
additional photons are found in an event, the combination of photons with the least $\chi^2$ is retained for further analysis.
The resulting $\chi^2$ of the kinematic fit is required to be less
than a decay mode dependent value, ranging from $25$ to $90$, which is obtained by optimizing the figure-of-merit
$N_S^{\rm MC}/\sqrt{N_S^{\rm data}+N_B^{\rm data}}$, where $N_S^{\rm MC}$ is the number of events from the
signal MC sample, and $N_S^{\rm data}$ and $N_B^{\rm data}$ represent the numbers of signal and background
events from data, respectively.

An inclusive MC sample containing $3.64\times10^{8}$ $\psi(3686)$ events and 48~pb$^{-1}$ of data collected at center-of-mass energy
$\sqrt{s} = 3.65$~GeV~\cite{condata}, which is about one fifteenth of the integrated
luminosity of the $\psi(3686)$ data, are employed to investigate the potential backgrounds.
Studies of the MC sample indicate the common backgrounds for all decay modes
are from $\psi(3686)\to \pi^0 +X$ ($X$ represents all possible
final states) and $\psi(3686)\to \pi^+\pi^- J/\psi$ decays. The former one is suppressed by requiring the invariant mass of any
two photons $M_{\gamma\gamma}$ to be out of the $\pi^0$ mass region, $|M_{\gamma\gamma}-M_{\pi^0}|>15$~MeV/$c^2$, where $M_{\pi^0}$ is
the nominal $\pi^0$ mass~\cite{pdg}. The latter one is suppressed by requiring the recoil
mass of any $\pi^+\pi^-$ combination $M^{\rm rec}_{\pi^+\pi^-}$ to be out of the $J/\psi$ mass region
$|M^{\rm rec}_{\pi^+\pi^-}-M_{J/\psi}|>5$~MeV/$c^2$, where $M_{J/\psi}$ is
the nominal $J/\psi$ mass~\cite{pdg}. For the $\chi_{c0,2}\to\eta\eta'$ channel, there is background
from $\chi_{c0,2}\to\gamma J/\psi, J/\psi \to \gamma \eta'$, which is suppressed
by further requiring the invariant mass of any $\gamma \eta'$ combination to be out of the
region (3.05, 3.16) GeV/$c^2$ for mode I and (3.049, 3.199) GeV/$c^2$ for mode II, respectively, where
the $\gamma$ is from the $\eta$ candidates.
The cross contaminations
between different decay modes are studied and are found to be negligible.
For the data at $\sqrt{s} = 3.65$~GeV,
there are almost no events satisfying the above selection criteria, which indicates
that the background due to continuum production is negligible.

\begin{figure}[hbtp]
\begin{minipage}{0.22\textwidth}
\includegraphics[width=\textwidth]{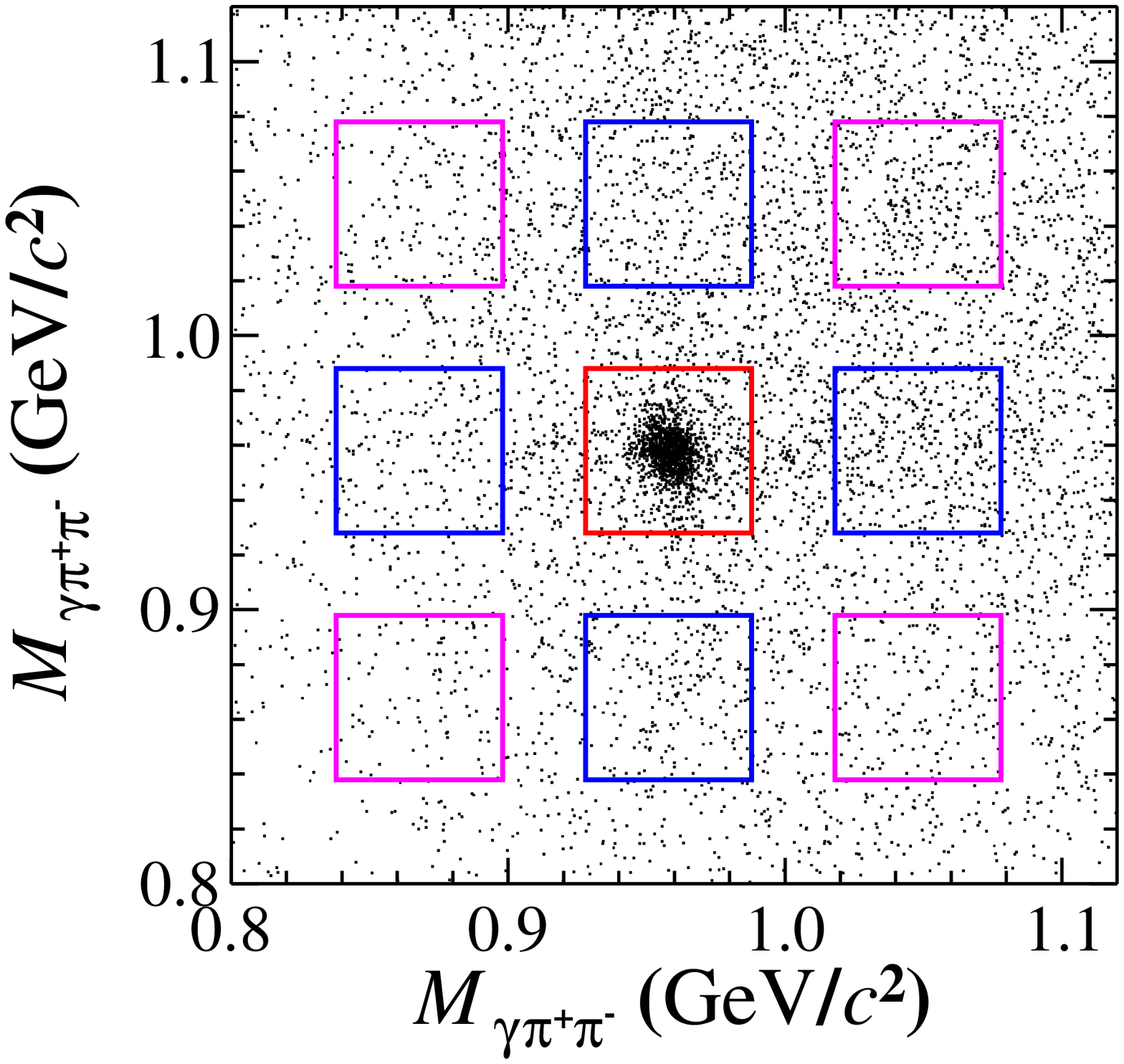}
\put(-80,95){ \bf (a)}
\end{minipage}
\begin{minipage}{0.22\textwidth}
\includegraphics[width=\textwidth]{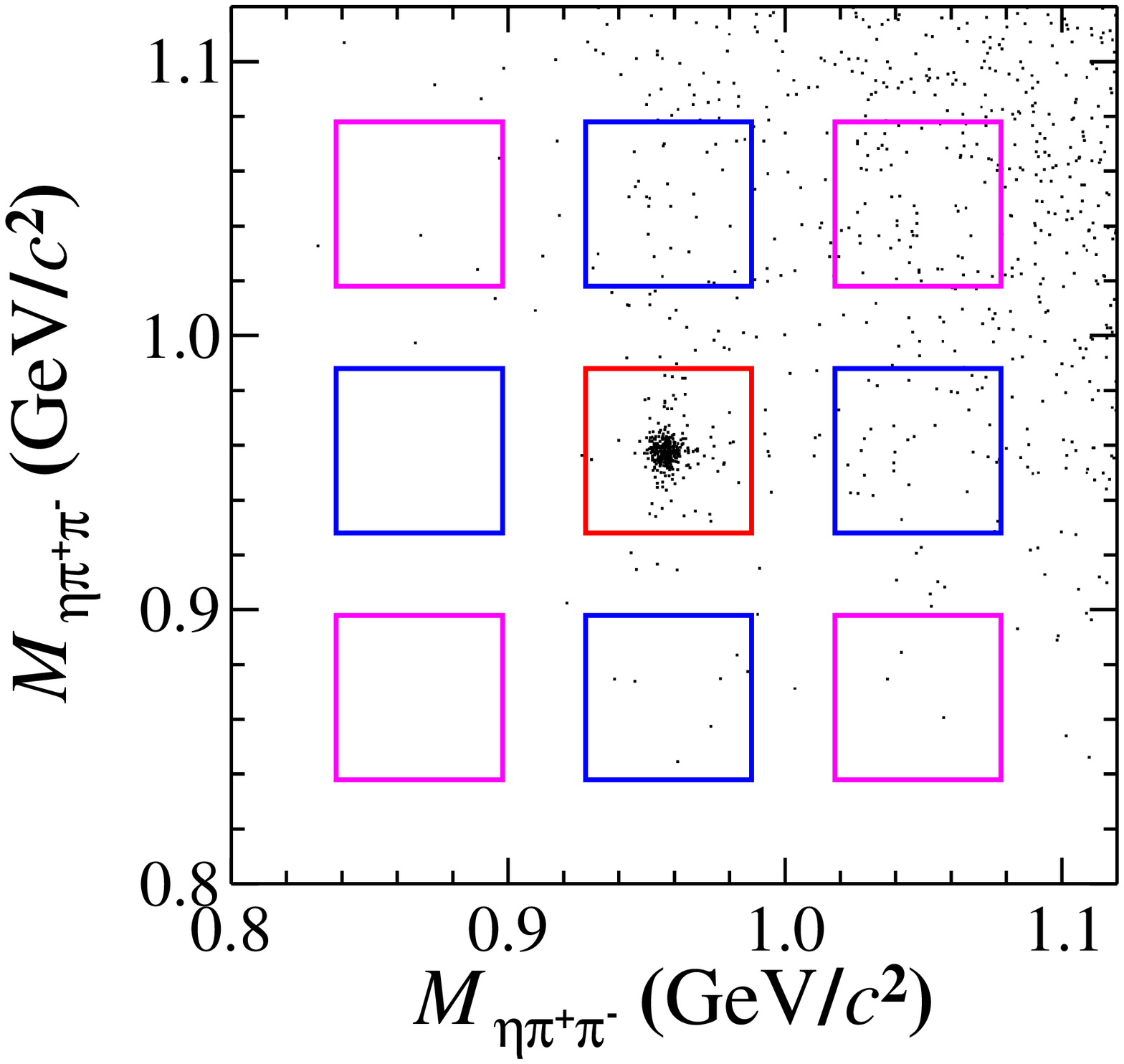}
\put(-80,95){ \bf (b)}
\end{minipage}
\vspace{1cm}
\begin{minipage}{0.22\textwidth}
\includegraphics[width=\textwidth]{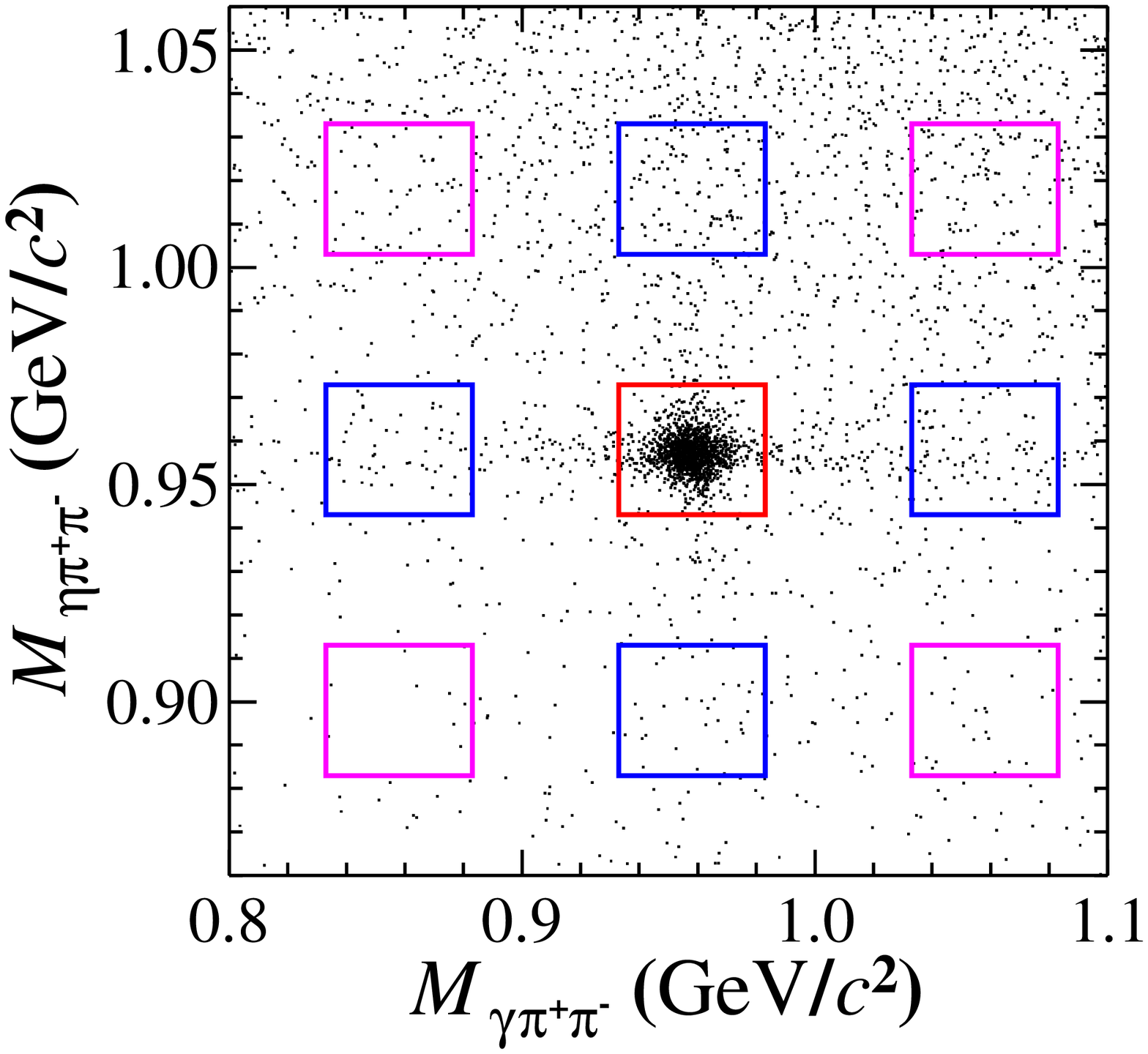}
\put(-80,95){ \bf (c)}
\end{minipage}
\caption{\small Scatter plots of $M_{i}$ versus $M_{j}$ of the candidate events for modes
(a) A, (b) B, and (c) C from the $\psi(3686)$ data. The boxes denote the signal and
background regions described in the text.}
\label{fig:scat}
\end{figure}

For the $\chi_{c0,2}\to\eta'\eta'$ decay, the two $\eta'$ candidates are selected by minimizing $(M_{i}-M_{\eta'})^{2}+(M_{j}-M_{\eta'})^{2}$.
Here, the subscripts $i/j = 1$ or $2$ denote $\gamma\pi^+\pi^-$ or $\eta\pi^+\pi^-$ for the two
different decay modes, respectively, and $M_{\eta'}$ is the $\eta'$ nominal mass~\cite{pdg}.
Figures~\ref{fig:scat}(a), (b) and (c) show the scatter plots of
$M_{i}$ versus $M_{j}$ of the candidate events for the modes A, B, and C individually.
The double-$\eta'$ signal region is defined as $M_1\in (0.943,0.973)$~GeV/$c^2$ for mode A,
$M_2 \in (0.928, 0.988)$~GeV/$c^2$ for mode B, and $M_1\in (0.933,0.983)$~GeV/$c^2$
and $M_2 \in (0.943, 0.973)$~GeV/$c^2$ for mode C.
Clear double-$\eta'$ signals are seen in the intersection region (shown as the central square) for each mode.
The eight squares with equal area around the signal region are selected to be sideband regions, which
are classified into two categories: the four boxes in the corners are used to estimate the
background contribution from background without $\eta'$ in subsequent decays (namely type A),
and the remaining four boxes are
used to estimate the background with one $\eta'$ in subsequent decays (namely type B).

\begin{figure}[hbtp]
\begin{minipage}{0.23\textwidth}
\includegraphics[width=\textwidth]{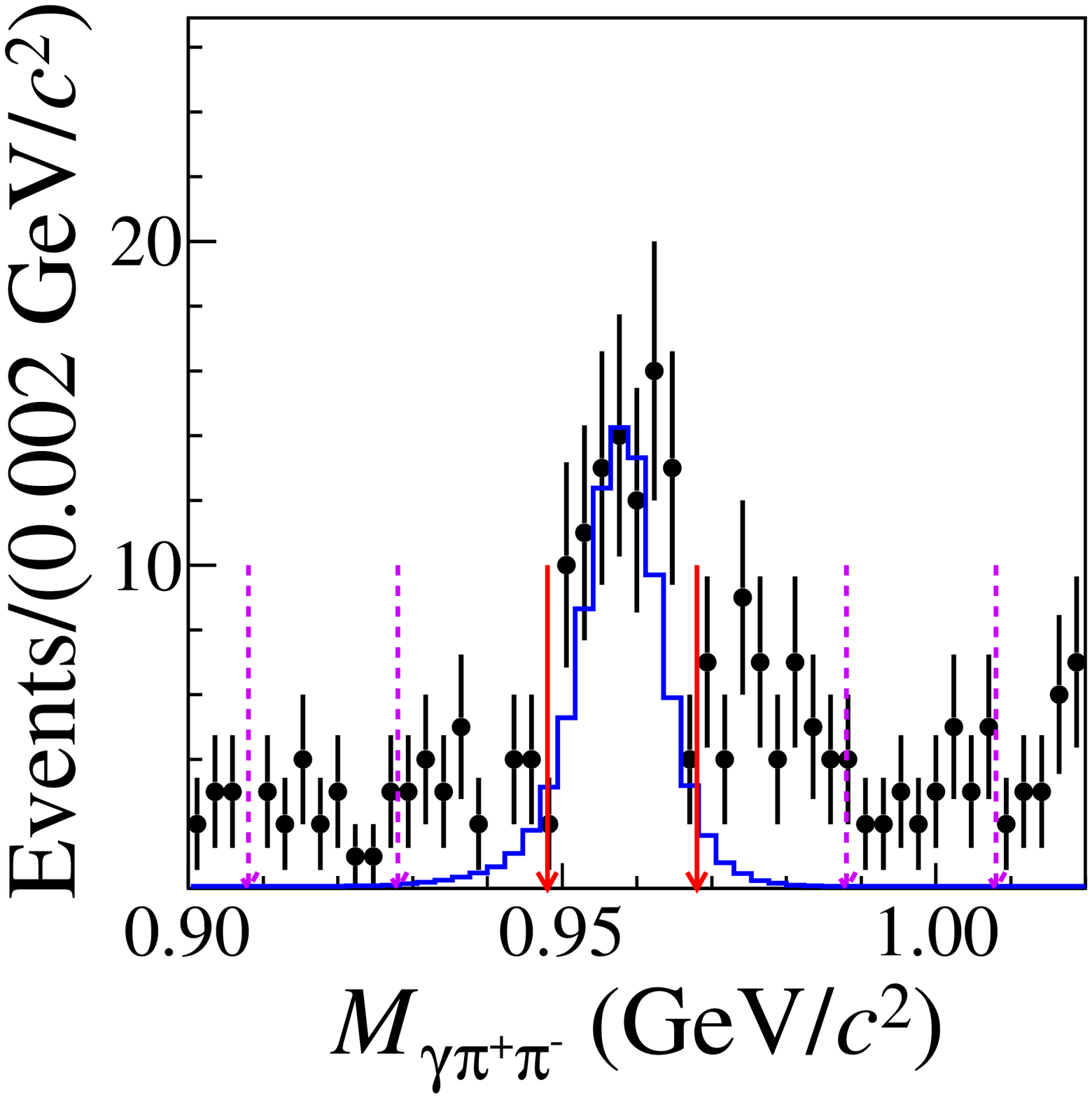}
\put(-80,90){\bf (i)}
\end{minipage}
\begin{minipage}{0.23\textwidth}
\includegraphics[width=\textwidth]{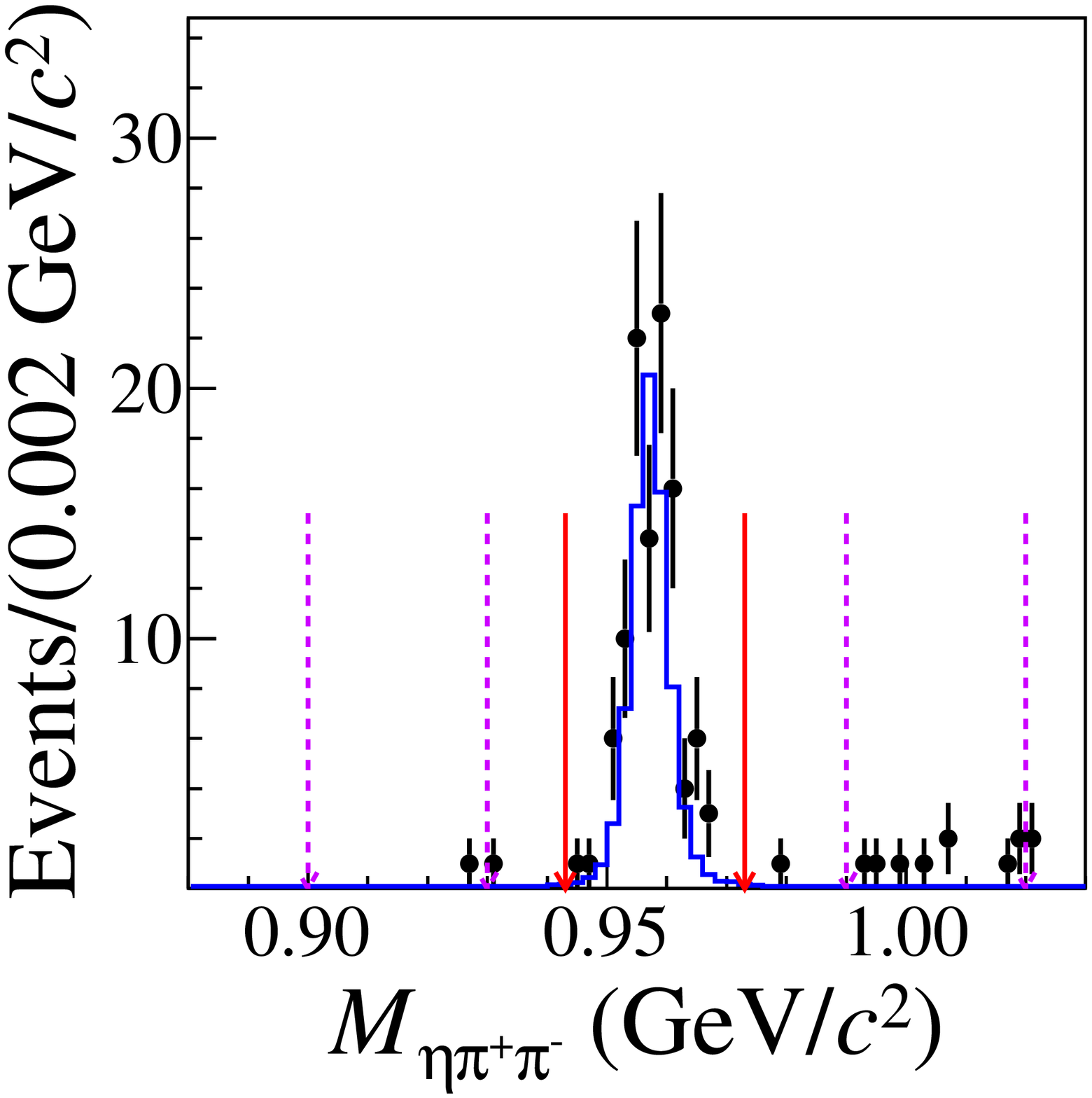}
\put(-80,90){\bf (ii)}
\end{minipage}
\caption{\small The $M_i$ distributions of the $\eta'$ candidate events for
modes (i) I and (ii) II.
In each plot, the dots with error bars are for the $\psi(3686)$ data,
and the histograms are for the signal MC samples, the solid arrows show the
$\eta'$ signal regions and the dashed ones show sideband regions.}
\label{fig:etap}
\end{figure}

For the $\chi_{c0,2}\to\eta\eta'$ decay, the $\eta'$ candidate is selected if it has a minimum $|M_i-M_{\eta'}|$.
Figure~\ref{fig:etap} shows the $M_{i}$ distributions of the candidate events for the two $\eta'$
decay modes, where clear $\eta'$ signals are observed in both modes. The $\eta'$ signal
region is defined as $M_{1}\in (0.948,0.968)$~GeV/$c^2$ or $M_{2} \in (0.943,0.973)$~GeV/$c^2$,
and two sideband regions with width equal to that of the signal region are chosen around the
signal region for each decay mode.

\section{data analysis}

\begin{figure}[hbtp]
\begin{center}
\begin{minipage}{0.49\textwidth}
\centering
\includegraphics[width=0.95\textwidth]{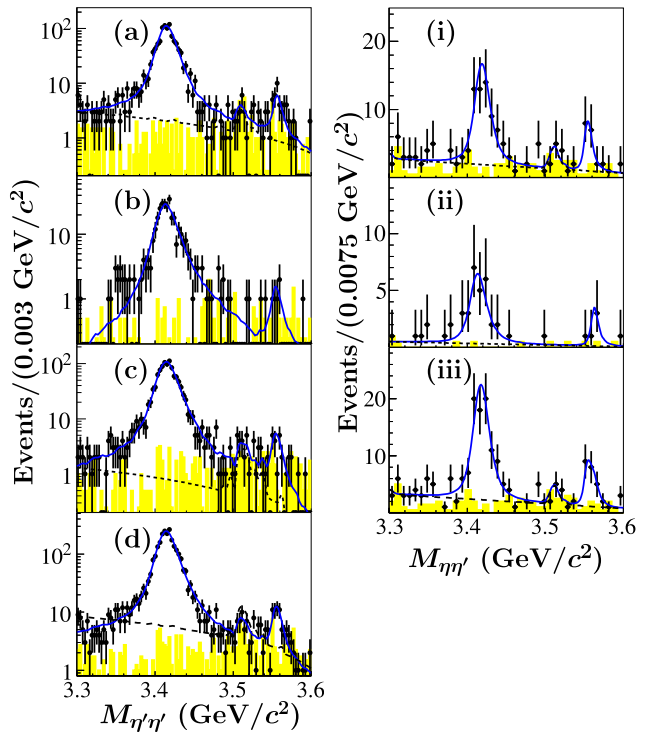}
\end{minipage}
\caption{\small Left column shows the simultaneous fits for $\chi_{c0,2}\to\eta'\eta'$.
(a) Mode A. (b) Mode B. (c) Mode C. (d) Sum of (a), (b), and (c).
Right column shows the simultaneous fits for $\chi_{c0,2}\to\eta\eta'$. (i) Mode I. (ii) Mode II. (iii) Sum of (i) and (ii).
In all of the above plots, the dots with error bars denote the $\psi(3686)$ data, the solid line denotes the overall fit results,
the dashed line denotes the backgrounds and the yellow histogram shows the normalized events in the $\eta'$ sideband regions. }
\label{fig:simfit}
\end{center}
\end{figure}

Figure~\ref{fig:simfit}(a)-(c) shows the spectra of $\eta'\eta'$ invariant mass
$M_{\eta'\eta'}$ for the candidate
events in the modes A, B, and C, respectively, while Fig.~\ref{fig:simfit}(d) shows the
corresponding distribution summed over the three decay modes.
Clear $\chi_{c0,2}$ signals are observed.
The expected background, which is estimated with the events within the sideband
regions normalized by $\frac{1}{2}M_{\rm side}^{\rm B} - \frac{1}{4}M_{\rm side}^{\rm A}$,
are presented as histograms in the corresponding figures, where $M_{\rm side}^{\rm A}$ and $M_{\rm side}^{\rm B}$
are the corresponding distributions in the sidebands A and B regions, and we assume the background is
distributed uniformly around the $\eta'$ signal region.
No obvious $\chi_{c0,2}$ peaks are found in the sideband regions, while $\chi_{c1}$ peaks are seen in modes A and C.
A study with the inclusive MC sample indicates
that the small bump around the $\chi_{c1}$ mass region for mode A comes from the
$\chi_{c1}\to\gamma J/\psi$, $J/\psi\to\gamma 2(\pi^+\pi^-)$ channel, while that for mode C comes from
$\chi_{c1}\to f_0(980)\eta'$,
 which will be considered later.

Figures~\ref{fig:simfit}(i) and (ii) show the distributions of $\eta\eta'$ invariant mass $M_{\eta\eta'}$
for the two $\eta'$ decay modes, where clear $\chi_{c0,2}$ signals are visible. The normalized
events in the $\eta'$ sideband region are also depicted and no obvious $\chi_{c0,2}$ peaks are
observed, while the $\chi_{c1}$ signal is seen in mode I. Analysis with an inclusive MC sample indicates that the small
$\chi_{c1}$ bump in mode I comes from the processes $\psi(3686)\to\gamma\chi_{c1}$, $\chi_{c1}\to\gamma J/\psi$,
$J/\psi\to \gamma\gamma\pi^+\pi^- (\eta\pi^+\pi^-$ or $\gamma\eta'$ with $\eta'\to\gamma\pi^+\pi^-$, etc.),
which will be taken into account in the fit later.

\vspace{0.2cm}

To determine the branching fractions of $\chi_{c0,2} \to \eta'\eta'$ and $\eta\eta'$,
two simultaneous fits to the three $M_{\eta'\eta'}$ spectra and the two $M_{\eta\eta'}$ spectra
are performed. The overall probability density functions in fitting
include three components: the $\chi_{c0,2}$ signals,
the $\chi_{c1}$ peaking background for specific modes, and the non-peaking
background. In the fit, the $\chi_{c0,2}$ signals are described with the MC-simulated shape of histogram convolved
with a Gaussian function to compensate for the potential resolution difference between data and MC simulation.
Due to limited-size of data sample, the parameters of the Gaussian function are fixed to those obtained from control samples,
such as $\psi(3686)\to\gamma\chi_{c0,2}$ with
$\chi_{c0,2}\to 2(\pi^+\pi^-)$, $\psi(3686)\to \gamma\chi_{c0,2}$ with $\chi_{c0,2}\to 2(\pi^+\pi^-\pi^0)$, which have
similar final states of interest. The shape of the $\chi_{c1}$ peaking background for the specific
modes are described with the MC simulation of the corresponding background modes, and their magnitudes are floated.
The non-peaking backgrounds are described by a first order Chebychev polynomial.
In the fit, the branching fractions of $\chi_{c0,2}\to\eta'\eta'/\eta\eta'$, ${\mathcal B}(\chi_{c0,2}\to\eta'\eta'/\eta\eta')$,
are taken as the common parameters among the different decay modes.
The projections of the simultaneous fit are shown in Fig.~\ref{fig:simfit}. The statistical
significance are $9.6 \sigma$ for $\chi_{c2}\to \eta'\eta'$,
$13.4\sigma$ for $\chi_{c0}\to \eta\eta'$ and $7.5\sigma$ for $\chi_{c2}\to \eta\eta'$, individually,
which are determined by comparing the fit likelihood values with and without the corresponding
$\chi_{c0,2}$ signal included.
The detection efficiencies $\epsilon$, the $\chi_{c0,2}$ signal yields in the different decay
modes, and the resultant decay branching fractions are summarized in Table~\ref{tab:results},
where the signal yields in each decay mode are calculated according to the total number $N_{\psi(3686)}$ of
$\psi(3686)$ events, the detection efficiency and the product
branching fractions in the subsequent decay. For mode C, there is a factor of two to account for the identical
particles. Except for the $\mathcal{B}(\chi_{c0,2}\to \eta'\eta'/\eta\eta')$
obtained in this measurement, all other decay branching fractions are taken from the PDG~\cite{pdg}.
The fitted numbers of $\chi_{c1}$ background are found to be consistent with the expectations from
the MC simulation.

\begin{table*}[ht]
\caption{\small The results for $\chi_{c0,2}\to \eta'\eta'/\eta\eta'$. $\mathcal B$ denotes branching fraction.}
\begin{center} {\scriptsize
\doublerulesep 2pt
\begin{tabular}{lcccccccccc} \hline
Decay channel            & \multicolumn{3}{c}{$\chi_{c0}\to\eta'\eta'$}        & \multicolumn{3}{c}{ $\chi_{c2}\to\eta'\eta'$}
 & \multicolumn{2}{c}{$\chi_{c0}\to\eta\eta'$}        & \multicolumn{2}{c}{ $\chi_{c2}\to\eta\eta'$}  \\\hline
$\eta'$ decay mode     & Mode A   & Mode B   & Mode C  & Mode A   & Mode B   & Mode C  & Mode I              & Mode II          & Mode I             & Mode II  \\
Efficiency(\%)    &$\phantom{1}12.9 \pm 0.1$    &$11.9 \pm 0.1$   &$\phantom{1}13.0 \pm 0.1$
                  &$14.0 \pm 0.1$    &$14.8 \pm 0.1$   &$14.9 \pm 0.1$
                  &$12.7 \pm 0.1$    &$\phantom{1}9.0 \pm 0.1$   &$14.7 \pm 0.1$    & $10.4 \pm 0.1$\\
Signal number     &$1057\pm 15$    &$329 \pm 5\phantom{1}$      &$1238 \pm 17$
                  &$22.7 \pm 2.6$    &$\phantom{1}8.1 \pm 0.9$      &$28.1 \pm 3.3$
                  &$59.9 \pm 5.3$    &$24.1 \pm 2.1$    &$14.3 \pm 2.8$    &$5.5 \pm 1.1$   \\
$\mathcal B$ (This work)  & \multicolumn{3}{c}{$(2.19\pm0.03\pm0.14)\times10^{-3}$}  & \multicolumn{3}{c}{$(4.76\pm0.56\pm0.38)\times10^{-5}$ }
                          & \multicolumn{2}{c}{$(8.92\pm0.84\pm0.65)\times 10^{-5}$}  & \multicolumn{2}{c}{$(2.27\pm0.43\pm0.25)\times 10^{-5}$}\\
$\mathcal B$ (PDG)~\cite{pdg} & \multicolumn{3}{c}{$\phantom{123456}(1.96\pm0.21)\times10^{-3}$}
                              & \multicolumn{3}{c}{$\phantom{123456789012}<1.0\times10^{-4}$}
                              & \multicolumn{2}{c}{$\phantom{1234567890123}<23\times10^{-5}$}
                              & \multicolumn{2}{c}{$\phantom{123456789012}<6.0\times10^{-5}$} \\
\hline
\end{tabular}
\label{tab:results}}
\end{center}
\end{table*}

\section{systematic uncertainty}

Several sources of systematic uncertainty in the branching fraction measurements are considered.
The systematic uncertainty from the total number of $\psi(3686)$ events, estimated by measuring inclusive
hadronic events, is $0.7$\%~\cite{npsip,npsip1}.
The uncertainty from MDC tracking and photon detection have been studied with the high purity control sample
of $\psi(3686)\to\pi^+\pi^-J/\psi$, $J/\psi\to l^{+}l^{-}$ and $J/\psi\to\rho\pi$.
The difference in the detection
efficiency between data and MC simulation is less than $1$\% per charged track, which is taken
as the systematic uncertainty~\cite{npsip}.
Employing a method similar to that in Ref.~\cite{c3}, except using a larger $J/\psi$ data set~\cite{njpsi2},
the difference of the photon detection efficiency between data and MC simulation is determined to be within
0.5\% in the barrel and 1.5\% in the endcaps of the EMC.
%RYAN: What is the larger J/psi data set?
In this analysis, the weighted uncertainty is 0.6\% per photon by considering the photon angular distribution.
The uncertainty due to $\eta$ reconstruction is determined by using a high purity control sample of
$J/\psi\to\eta p\bar{p}$ decays. The difference of $\eta$ reconstruction efficiencies between data and MC simulation,
about $1.0$\% per $\eta$~\cite{c1}, is taken as the systematic uncertainty.
The uncertainty from the $\eta'$ mass window requirement is estimated
by changing the $\eta'$ signal windows by one unit of the mass resolution.
The resultant difference in the branching fractions is taken as the systematic uncertainty.
The uncertainty related to the kinematic fit is due to the inconsistency between data and MC simulation of the
track parameters and their error matrices. In this work,
only charged pions are involved and their track parameters in MC simulation
are corrected by using the control sample
$\psi(3686)\to\pi^+\pi^-K^{+}K^{-}$. As a consequence, the consistency between
data and MC simulation is significantly improved.
The difference of the detection efficiencies with and without the correction is taken as the
uncertainty due to the kinematic fit. The detailed method to estimate the uncertainty of the
kinematic fit can be found in Ref.~\cite{c4}.
The uncertainty in the fit arises from resolution compensation, fit range and
background shape. The resolution compensation uncertainty is obtained by changing the
width of Gaussian function to the most conservative value estimated by the different control samples.
The uncertainties from fit range and background shape are estimated by shifting up or down the fit
intervals by 10~MeV/$c^2$ and by changing the order of the Chebychev polynomial function, respectively.
Summing the maximum uncertainties of each aspect in quadrature yields the uncertainty from the fit.
The uncertainty from decay branching fractions of intermediate states
in the subsequent decays is determined by setting the branching fractions, ${\cal B}(\psi(3686)\to\gamma\chi_{cJ})$,
${\cal B}(\eta'\to \gamma\pi^+\pi^-)$, ${\cal B}(\eta'\to \eta\pi^+\pi^-)$, and ${\cal B}(\eta\to\gamma\gamma)$,
randomly according to the Gaussian distributions, where the means and standard deviations of Gaussian
functions are taken to be their central values of the branching fractions and the corresponding
uncertainties in the PDG~\cite{pdg}. We repeat the same fitting process 100 times, and the standard
deviations of the resultant branching fractions are taken as the systematic uncertainty.
The uncertainty arising from the $\psi(3686)\to\pi^0 +X$ background subtraction is
estimated by changing the $\pi^0$ mass window $|M_{\gamma\gamma}-M_{\pi^0}|$ by $\pm1$~MeV/$c^2$ in
the event selection. Similarly, the uncertainty related to
$\psi(3686)\to\pi^+\pi^- J/\psi$ is estimated by changing the $J/\psi$ mass window
$|M_{\pi^+\pi^-}^{\rm recoil}-M_{J/\psi}|$ by $1$~MeV/$c^2$. The uncertainty arising from the
veto $\chi_{c0,2}\to \gamma J/\psi$ with $J/\psi \to \gamma \eta'$ is estimated by shifting
the $J/\psi$ mass window by $\pm1$ MeV/$c^2$.

Table~\ref{tab:sumerr} summarizes all the systematic uncertainties for $\chi_{c0,2}\to\eta'\eta'$
and $\chi_{c0,2}\to\eta\eta'$, in which the uncertainties from photon efficiency, $\eta$ reconstruction,
kinematic fit, and background veto are decay mode dependent, and the weighted average uncertainties are presented.
The weights are the product of the detection efficiency and the branching fractions of $\eta'$
and $\eta$ subsequent decays in individual decay modes.
The total systematic uncertainty is obtained by adding all individual values in quadrature.

\begin{table}[htbp]
\caption{\small The systematic uncertainties (in \%) in the branching fraction measurement.}
\begin{center}
\begin{tabular}{lcccc}\hline
\multirow{2}{*}{Decay channel}    &  \multicolumn{2}{c}{$\chi_{c0} \to $}  &\multicolumn{2}{c}{$\chi_{c2} \to $}   \\
 \multicolumn{1}{l}{}                                 &~~~~$\eta'\eta'$~~~~&~~~~$\eta\eta'$~~~~&~~~~$\eta'\eta'$~~~~  &~~~~$\eta\eta'$~~~~  \\   \hline
$N_{\psi(3686)}$                  & $0.6$         & $0.6$     & $0.6$    &  $0.6$  \\
Tracking                          & $4.0$         & $2.0$     & $4.0$    &  $2.0$  \\
Photon efficiency                 & $2.2$         & $2.6$     & $2.2$    &  $2.6$  \\
$\eta$ reconstruction             & $0.7$         & $1.3$     & $0.7$    &  $1.3$  \\
$\eta'$ mass window               & $1.0$         & $1.7$     & $1.0$    & $1.7$  \\
Kinematic fit                     & $0.7$         & $1.0$     & $0.6$    &  $1.7$  \\
%$\chi_{c0,2}$ signal fitting      & $1.7$         & $5.1$     & $2.3$    &  $5.5$  \\
$\chi_{c0,2}$ signal fitting      & $1.1$         & $5.0$     & $3.9$    &  $9.5$  \\
Intermediate state ${\mathcal B}$ & $3.8$         & $3.1$     & $4.4$    &  $3.8$  \\
Veto $\pi^+\pi^-J/\psi$           & $0.1$         &  -        & $0.9$    &  -    \\
Veto $\psi(3686) \to \pi^0+X$     & $0.2$         & $1.0$     & $2.1$    &  $0.2$  \\
Veto $J/\psi\to\gamma\eta'$       & -             & $0.8$     & -        &  $1.5$  \\ \hline
Total                             & $6.3$         & $7.3$     & $8.0$    &  $11.2$  \\
\hline
\end{tabular}
\label{tab:sumerr}
\end{center}
\end{table}

\section{summary}

In summary, based on $448.1\times 10^6$  $\psi(3686)$ events collected with the BESIII detector,
the decays $\chi_{c2}\to\eta'\eta'$, $\chi_{c0}\to\eta\eta'$ and $\chi_{c2}\to\eta\eta'$ are observed
for the first time with significances of $9.6\sigma$, $13.4\sigma$ and $7.5\sigma$, respectively,
and the corresponding branching fractions are measured.
The branching fraction of the decay $\chi_{c0}\to\eta'\eta'$ is also measured with improved precision. Table~\ref{tab:results}
summarizes the measured branching fractions of $\chi_{c0,2}\to\eta'\eta'$ and $\eta\eta'$.
With the measured
branching fractions, the relative strength $r$ between the DOZI and SOZI violating amplitudes for the
$\chi_{c0}$ and $\chi_{c2}$ decays to $PP$ final states, is estimated to be around $-0.15$ according
to Eq.~($15$) in Ref.~\cite{zhaoq1} with its input parameters.
This implies that the contribution from the DOZI violating amplitude is suppressed in $\chi_{c0,2} \to PP$
decays in comparison with the SOZI ones~\cite{zhaoq1,zhaoq2}.
In addition, we find ${\mathcal B}(\chi_{c0}\to\eta'\eta')/{\mathcal B}(\chi_{c2}\to\eta'\eta') \approx 45$,
which is about one order larger than the ratios for other pseudoscalar meson pairs, ranging from $3$ to $6$ for
$\pi^+\pi^-$, $\pi^0\pi^0$, $K^+K^-$, $K_S^0K_S^0$, $\eta\eta$~\cite{pdg} and $\eta\eta'$.
This large ratio is expected by the model proposed in Ref.~\cite{zhaoq1} given a relatively suppressed
DOZI-violating contribution. This may initiate further studies about the dynamics of $\chi_{c0,2} \to PP$.

The BESIII collaboration thanks the staff of BEPCII and the IHEP computing
center for their strong support. This work is supported in part by the National
Key Basic Research Program of China under Contract No. 2015CB856700; National
Natural Science Foundation of China (NSFC) under Contracts Nos. 11575077,
11475090, 11475207, 11605042, 11235011, 11322544, 11335008, 11425524, 11305090, 11235005, 11275266;
The China Scholarship Council; The Innovation Group of Nuclear and Particle Physics in USC;
the Chinese Academy of Sciences (CAS) Large-Scale Scientific Facility Program;
the CAS Center for Excellence in Particle Physics (CCEPP); the Collaborative
Innovation Center for Particles and Interactions (CICPI); Joint Large-Scale
Scientific Facility Funds of the NSFC and CAS under Contracts Nos. U1232201,
U1332201, U1532257, U1532258; CAS under Contracts Nos. KJCX2-YW-N29, KJCX2-YW-N45;
100 Talents Program of CAS; National 1000 Talents Program of China; INPAC and
Shanghai Key Laboratory for Particle Physics and Cosmology; German Research
Foundation DFG under Contracts Nos. Collaborative Research Center CRC 1044,
FOR 2359; Istituto Nazionale di Fisica Nucleare, Italy; Koninklijke Nederlandse
Akademie van Wetenschappen (KNAW) under Contract No. 530-4CDP03; Ministry of
Development of Turkey under Contract No. DPT2006K-120470; National Science and
Technology fund; The Swedish Resarch Council;
U. S. Department of Energy under Contracts Nos. DE-FG02-05ER41374, DE-SC-0010504,
DE-SC0012069; University of Groningen (RuG) and the Helmholtzzentrum f\"{u}r
Schwerionenforschung GmbH (GSI), Darmstadt; WCU Program of National Research
Foundation of Korea under Contract No. R32-2008-000-10155-0


\begin{thebibliography}{99}
\bibitem{zhaozou} Q.~Zhao, B.~S.~Zou and Z.~B.~Ma, \ Phys. \ Lett. \ B {\bf 631}, 22 (2005).
\bibitem{pdg} C.~Patrignani {\it et al.} (Particle Data Group), Chin. \ Phys. \ C {\bf 40}, 100001 (2016).
\bibitem{zhaoq1} Q.~Zhao, \ Phys. \ Rev. \ D {\bf 72}, 074001 (2005).
\bibitem{zhaoq2} Q.~Zhao, \ Phys. \ Lett. \ B {\bf 659}, 221 (2008).
\bibitem{chicss} M.~Ablikim {\it et al.} (BES Collaboration), \ Phys. \ Rev. \ D {\bf 70}, 092002 (2004); \ Phys. \ Rev. \ D {\bf 72}, 092002 (2005).
\bibitem{vv} M.~Ablikim {\it et al.} (BESIII Collaboration), \ Phys. \ Rev. \ Lett. {\bf 107}, 092001 (2011).
\bibitem{npsip} M.~Ablikim {\it et al.} (BES Collaboration), Chin. \ Phys. \ C {\bf 37}, 063001 (2013).
\bibitem{npsip1} The total number of $\psi(3686)$ events taken at 2009 and 2012 is obtained
based on the same method in Ref.~\cite{npsip}. The preliminary number is determined to be $448.1\times 10^{6}$ with uncertainties
of 0.7\%.
\bibitem{bes3} M.~Ablikim {\it et al.} (BES Collaboration), Nucl. \ Instrum. \
Meth. \ A {\bf 614}, 345 (2010).
\bibitem{geant4} S.~Agostinelli {\it et al.} ({\sc GEANT4} Collaboration), Nucl. \ Instrum. \
Meth. \ A {\bf 506}, 250 (2003); J.~Allison {\it et al.}, IEEE Trans. Nucl. Sci. {\bf 53}, 270 (2006).
\bibitem{boost} Z.~Y.~Deng {\it et al.}, High Energy Physics \& Nuclear Physics {\bf 30}, 371 (2006).
\bibitem{kkmc} S.~Jadach, B.~F.~L.~Ward, and Z.~Was, Comput.\ Phys.\ Commun. {\bf 130}, 260 (2000); Phys. \ Rev. \ D {\bf 63}, 113009 (2001).
\bibitem{evtgen} R.~G.~Ping, Chin. \ Phys. \ C {\bf 32}, 599 (2008); D.~J.~Lange, Nucl. \ Instr. \ Meth. \ A {\bf 462}, 152 (2001).
\bibitem{lundcharm} J.~C.~Chen, G.~S.~Huang, X.~R.~Qi, D.H.~Zhang, Y.S.~Zhu, Phys. \ Rev. \ D {\bf 62}, 034003 (2000);
R.~L.~Yang, R.~G.~Rong, D.~Chen, Chin. \ Phys. \ Lett. {\bf 31}, 061301 (2014).
\bibitem{E1} E.~Eichten, K.~Gottfried, T.~Kinoshita, K.D.~Lane, T.M.~Yan, Phys. \ Rev. \ D {\bf 21}, 203 (1980).
\bibitem{liuzq} M.~Ablikim {\it et al.} (BESIII Collaboration), Phys. \ Rev. \ D {\bf 84}, 092006 (2011).
\bibitem{condata} M.~Ablikim {\it et al.} (BESIII Collaboration), Chin. \ Phys. \ C {\bf 37}, 123001 (2013).
\bibitem{c3} M.~Ablikim {\it et al.} (BESIII Collaboration), Phys. \ Rev. \ D {\bf 81}, 052005 (2010).
\bibitem{njpsi2} M.~Ablikim {\it et al.} (BESIII Collaboration), Chin. \ Phys. \ C {\bf 41}, 013001 (2017).
\bibitem{c1} M.~Ablikim {\it et al.} (BESIII Collaboration), Phys. \ Rev. \ Lett. {\bf 105}, 261801 (2010).
\bibitem{c4} M.~Ablikim {\it et al.} (BESIII Collaboration), Phys. \ Rev. \ D {\bf 87}, 012002 (2013).

\end{thebibliography}
\end{document}